\documentclass[prd,twocolumn,noeprint]{revtex4-1}
\usepackage{graphicx,natbib}
\usepackage{url}
\usepackage{color}
\usepackage{rotating}
\usepackage{amssymb,amsmath}

\def\hmpcinv{h\,{\rm Mpc}^{-1}}

\newcommand{\fnl}{f_{\rm NL}}
\newcommand{\fnlstar}{f_{\rm NL}^*}

\newcommand{\fsky}{f_{\rm sky}}

\newcommand{\mnras}{{Mon.~Not.~R.~Astron.~Soc.}}

\newcommand{\ellmax}  {\ell_{\mathrm{max}}}

\newcommand{\nf}{{n_{f_{\rm NL}}}}
\newcommand{\nhat}{\mathbf{\hat{n}}}
\newcommand{\perm}{{\rm perm.}}
\newcommand{\ells}[0]{\ell_1 \ell_2 \ell_3}
\newcommand{\fnlstarhat}{\hat{f}^*_{\rm NL}}
\newcommand{\kpiv}{k_{\rm piv}}

\begin{document}

\title{First constraints on the running of non-Gaussianity}

\author{Adam Becker}
\email{beckeram@umich.edu}
\affiliation{Department of Physics, University of Michigan, 
450 Church St, Ann Arbor, MI 48109-1040}

\author{Dragan Huterer}
\email{huterer@umich.edu}
\affiliation{Department of Physics, University of Michigan, 
450 Church St, Ann Arbor, MI 48109-1040}

\begin{abstract} 
We use data from the WMAP temperature maps to constrain a scale-dependent
generalization of the popular `local' model for primordial non-Gaussianity.
In the model where the parameter $\fnl$ is allowed to run with scale $k$,
$\fnl (k)= \fnlstar (k/\kpiv)^\nf$, we constrain the running to be
%\adam{Updated!} 
%$\nf=3.1^{+6.1}_{-3.0}$
$\nf =0.30^{+1.9}_{-1.2} $ at 95\% confidence, marginalized over the amplitude
$\fnlstar$. The constraints depend somewhat on the prior probabilities
assigned to the two parameters. In the near future, constraints from a
combination of Planck and large-scale structure surveys are expected to
improve this limit by about an order of magnitude and usefully constrain
classes of inflationary models.
\end{abstract}

\date{\today}

\maketitle

{\it Introduction.}  Non-Gaussianity in the distribution of primordial density
fluctuations provides a unique window into the physics of
inflation. The magnitude of primordial non-Gaussianity and its dependence on
scale provide information about the dynamics of scalar field(s), their
interactions, and the speed of sound during inflation. Constraints on
non-Gaussianity have traditionally come from the measurements of the
three-point correlation function of the cosmic microwave background (CMB)
temperature anisotropies. Upper limits from COBE \cite{COBE_fnl} have been
improved by two orders of magnitude by the WMAP experiment
\cite{wmap7}. Moreover, clustering of galaxies and galaxy clusters has also
been identified as a powerful probe of non-Gaussianity \cite{Dalal}, already
leading to interesting constraints that are complementary in their information
content to the CMB measurements.

So far most attention has been devoted to the ``local" model of primordial
non-Gaussianity, where the primordial Newtonian potential $\phi(x)$ is
modified with a quadratic term: $\phi = \phi_G + \fnl(\phi_G^2 -
\langle\phi_G^2\rangle )$, where $\phi_G$ is a Gaussian potential
\cite{Komatsu_Spergel}. The parameter $\fnl$ is currently constrained to be
$32\pm 21$ by WMAP (\cite{wmap7}; see also
\cite{Yadav_Wandelt_evidence,Smith_Sen_Zal_optimal}) and $28\pm 23$ by the
large-scale structure \cite{Slosar_etal,Afshordi_Tolley,Xia2011}.  Several
other non-Gaussian models have been constrained as well
(e.g.\ \cite{Fergusson2010,Smidt2010}).  However, the `running' with physical
scale of these models, which may carry important information about the
number of inflationary fields and their interactions
\cite{Chen2005,Liguori2006,Khoury_Piazza,Kumar2010,bryb,wandsb,Byrnes2010,chris5,
  Riotto2010,Barnaby:2012tk,Kobayashi_Takahashi}, has not yet been constrained
with current data (except for a very rough estimate of the angular-multipole
dependence of $\fnl$ \cite{Smidt2010} and implicit constraints on a
braneworld-motivated model \cite{Bean_DBI}). Such constraints have only been
forecasted for future experiments
\cite{Sefusatti2009,Becker2011,Shandera2010,Giannantonio_forecasts,Becker2012}.
Constraining the running of non-Gaussianity therefore presents a major new
opportunity to probe inflationary physics, and is just becoming feasible. In
this Letter, we present the first such constraints.

\medskip
{\it Model.}  In this work we consider a physically motivated generalization
of the local model, where the parameter $\fnl$ is promoted to a function of
scale $k$.  In particular, we seek to constrain the two-parameter power-law
subclass of the generalized models \cite{Becker2011}
\begin{equation}
\fnl(k) = \fnlstar \left ({k\over k_*}\right )^\nf ,
\label{eq:fnlk}
\end{equation}
where $k_*$ is an arbitrary fixed parameter, leaving $\fnlstar$ and $\nf$ as
the parameters of interest in this model. Such scaling is expected in
inflation when more than one field dominates or when there is
self-interaction, and its signatures in the CMB and LSS have been discussed in
the literature \cite{LoVerde,Sefusatti2009,Becker2011}. The parameter $\nf$ is
often, though certainly not always, expected to be $\lesssim O(1)$ in
inflationary models, but in our phenomenological model it is allowed to take
any value.

\medskip
{\it Bispectrum and $\fnlstar$ estimator.}
The primordial bispectrum of the $\fnl(k)$ model from Eq.~(\ref{eq:fnlk}) is
straightforward to calculate:
\begin{equation}
F(\vec{k}_1, \vec{k}_2, \vec{k}_3) = 2 \left [\fnl(k_1) P(k_2) P(k_3) +
  \perm\right ],
\end{equation}
where the full bispectrum is $B(\vec{k}_1,\vec{k}_2,\vec{k}_3)\equiv
(2\pi)^3\delta(\vec{k}_1+\vec{k}_2+\vec{k}_3)F(\vec{k}_1,\vec{k}_2,\vec{k}_3)$. Here
$P$ is the power spectrum of the primordial curvature perturbations, and $\delta$
is the Dirac delta function.

Constraining the running parameter $\nf$ seems difficult because of the
apparent requirement to find an estimator for a parameter in an exponent. To
avoid this, we resort to an indirect approach where, for a {\it fixed} value
of $\nf$, we estimate the parameter $\fnlstar$ using modifications of the
well-known KSW estimator \cite{KSW}, which is
  known to be nearly optimal
  \cite{Smith_Zaldarriaga,Creminelli_estimators}. We then iterate over the
values of the running $\nf$ to obtain the full likelihood
$\mathcal{L}(\fnlstar, \nf)$.

The theoretical expectation for the bispectrum of the temperature anisotropies
in the cosmic microwave background can be explicitly evaluated, starting from
the definition of the generalized non-Gaussian local model in Eq.~(\ref{eq:fnlk})
to account for the running $\nf$:
\begin{align}
B^{\rm theory}_{\ell_1 \ell_2 \ell_3} &(\fnl^*, \nf) = 2 \fnl^* I_{\ells}
\times \nonumber \\[0.2cm]
&\int_0^{\infty} r^2 d r \left( \alpha_{\ell_1}(\nf, r) \beta_{\ell_2}(r) 
\beta_{\ell_3}(r) + \mbox{perm.} \right)
\end{align}
where $I_{\ells}$ is the Gaunt integral and
\begin{eqnarray}
\alpha_{\ell}(r) &\equiv&  {
2 \over \pi} {1 \over k_{\rm piv}^{\nf}}  \int k^{2 + \nf} \: t_{\ell}(k)
j_{\ell}(kr) dk\\[0.2cm]
\beta_{\ell}(r) & \equiv & 
{2 \over \pi} \int k^2 P_\Phi (k) t_{\ell}(k) j_{\ell}(kr) dk. 
\label{eq:alpha_and_beta}
\end{eqnarray}
Here, $t_\ell$ is the radiation transfer function, which can be calculated
using CAMB \cite{CAMB}.
Following 
%Refs.~\cite{KSW,Smith_Zaldarriaga,Creminelli_estimators}
KSW \cite{KSW} we can
define new, filtered maps $A(\nhat, r)$ and $B(\nhat, r)$,
\begin{eqnarray}
\label{eq:A_map}
A(\nhat, r) &\equiv &
\sum_{\ell, m} \alpha_\ell (\nf, r) {b_\ell \over 
\tilde{C}_\ell} a_{\ell m} Y_{\ell m} (\nhat), \\[0.2cm]
B(\nhat, r) &\equiv &\sum_{\ell, m} \beta_\ell (r) 
{b_\ell \over \tilde{C}_\ell} a_{\ell m} Y_{\ell m} (\nhat).
\label{eq:B_map}
\end{eqnarray}
Then, we write down the skewness $S(\nf)$:
\begin{equation}
S(\nf) \equiv \int r^2 dr \int d^2 \nhat \,A(\nhat, r) B^2 (\nhat, r),
\label{eq:skew}
\end{equation}
which requires $\nf$ as input (through $A$), and does not require {\it a priori}
knowledge of $\fnlstar$.

The observed CMB bispectrum is defined as $B_{\ells}^{\rm obs.} = \langle
a_{\ell_1 m_1} a_{\ell_2 m_2} a_{\ell_3 m_3} \rangle$, and $S(\nf)$ therefore
reduces to
\begin{equation}
S = \sum_{\ell_1 \leq \ell_2 \leq \ell_3} {B^{\rm obs}_{\ells} 
\tilde{B}^{\rm theory}_{\ells}( \fnl = 1) \over \tilde{C}_{\ell_1} \tilde{C}_{\ell_2} \tilde{C}_{\ell_3} },
\label{SkewBispec}
\end{equation}
where $\tilde{B}^{\rm theory}_{\ells} = b_{\ell_1} b_{\ell_2} b_{\ell_3}
B^{\rm theory}_{\ells}$, and $b_\ell$ is the beam transfer function. 

We now define $F\equiv F(\nf)$, the Fisher matrix for $\fnlstar$, equivalent to the cumulative
signal-to-noise squared of the theoretical bispectrum for $\fnlstar=1$
\begin{equation}
F(\nf) = \sum_{\ell_1 \leq \ell_2 \leq \ell_3} 
{ \left(\tilde{B}^{\rm theory}_{\ells} (\fnlstar = 1) \right)^2 \over 
\tilde{C}_{\ell_1} \tilde{C}_{\ell_2} \tilde{C}_{\ell_3} }.
\label{eq:Fisher}
\end{equation}
The theoretical expectation for $B_{\ells} \propto \fnlstar$, so the cubic KSW
estimator for $\fnlstar$ is:
\begin{equation}
%\hat{f}_{\rm NL} = {S/\fsky + S_{\rm linear}\over F}.
%\hat{f}_{\rm NL} = {S\over F}.
\fnlstarhat = {S\over F}.
\label{eq:KSW}
\end{equation}

We used HEALPix, by way of HealPy, to do the forwards and backwards spherical
harmonic transforms required to obtain the $A$ and $B$ maps.

\medskip
{\it Cut-sky maps.}  Equation \eqref{eq:KSW} works well for a full-sky map,
but a sky cut introduces a spurious non-Gaussian signal.  To account for the
masking of the CMB sky, we make the substitution $S \rightarrow S_{\rm cut} =
S/\fsky + S_{\rm linear}$ \cite{Yadav2008}. $S_{\rm linear}$ is an addition to
skewness from Eq.~(\ref{eq:skew}), calibrated to account for partial-sky
observations:
\begin{align}
S_{\rm linear}  = & - {1 \over \fsky} \int r^2 dr \int d^2 \nhat  
\left[ A(\nhat, r) \langle B_{\rm sim}^2 (\nhat, r) \rangle_{MC} \right. \nonumber
 \\
& \left. + 2B(\nhat, r) \langle A_{\rm sim}(\nhat, r) B_{\rm sim}(\nhat, r) \rangle_{MC} \right].
\end{align}
The subscripted filtered maps $A_{\rm sim}$ and $B_{\rm sim}$ are created
from Python-produced 
Monte Carlo realizations of the cut CMB sky; the brackets $\langle
\rangle_{MC}$ indicate an average over all 300 Monte-Python maps. The simulated
maps were produced using the prescription laid out in Appendix A of the WMAP5
paper \cite{wmap5}; the only difference (aside from using the WMAP7
cosmological model) is that we used a uniform weighting for the maps,
rather than the slightly more complicated weighting given there, since it only
gives a marginal improvement in estimating $\fnl$.

\medskip
{\it Likelihood Evaluation.}  To find the likelihood, we first find a $\chi^2$
statistic for $\fnl^*$, given a value of $\nf$.  Taking the angular-averaged
bispectrum $B_{\ells}$ as our observables, we have:
\begin{align}
\chi^2 &(\fnl^*, \nf) 
= 
\nonumber \\
&\sum_{\ells} {\left( B^{\rm obs}_{\ells} - \fnl^* 
\tilde{B}^{\rm theory}_{\ells}(\nf, \fnl^* = 1) \right)^2 \over \tilde{C}_{\ell_1} \tilde{C}_{\ell_2} \tilde{C}_{\ell_3} } 
\end{align}
(Again, this works because the theoretical expectation for $B_{\ells} \propto
\fnlstar$.)  Using Eqs.~\eqref{SkewBispec} and \eqref{eq:Fisher}, we can
rewrite $\chi^2$ as
\begin{equation}
\chi^2(\fnl^*, \nf) =  F \left( \fnl^* - {S \over F} \right)^2 + \chi^2_0 - {S^2 \over F}.
\label{eq:chi2}
\end{equation}
where $\chi^2_0 \equiv \sum_{\ells} \left(B^{\rm obs}_{\ells} \right)^2/
(\tilde{C}_{\ell_1} \tilde{C}_{\ell_2} \tilde{C}_{\ell_3})$ is the
goodness-of-fit parameter for the data with respect to the $\fnl^* = 0$
case. Note that the numerator of $\chi^2_0$ is an observed quantity, and the
denominator is based solely on the theoretical prediction for the power
spectrum (as well as a few noise and beam parameters of WMAP). Therefore,
$\chi^2_0$ does not depend on $\fnl^*$ or $\nf$ at all.  We can use
the definition of $\fnlstarhat$ in Eq.~\eqref{eq:KSW} to rewrite the
expression for $\chi^2$ as follows
\begin{equation}
\chi^2(\fnl^*, \nf) =  F \left( \fnl^* -  \hat{f}^*_{\rm NL} \right)^2 + \chi^2_0 - (\fnlstarhat)^2 F.
\end{equation}
For a fixed value of $\nf$, the $\chi^2$ is, as expected, minimized in
$\fnl^*$ when $\fnl^* = \fnlstarhat$, and one obtains $\chi_{\rm min}^2(\nf) =
\chi^2_0 - (\fnlstarhat)^2 F$.

\begin{figure*}[t] %htbp, h = here, t = top, b = bottom, p = page: if nothing there...
\begin{center}
\includegraphics[width= 3.5in]{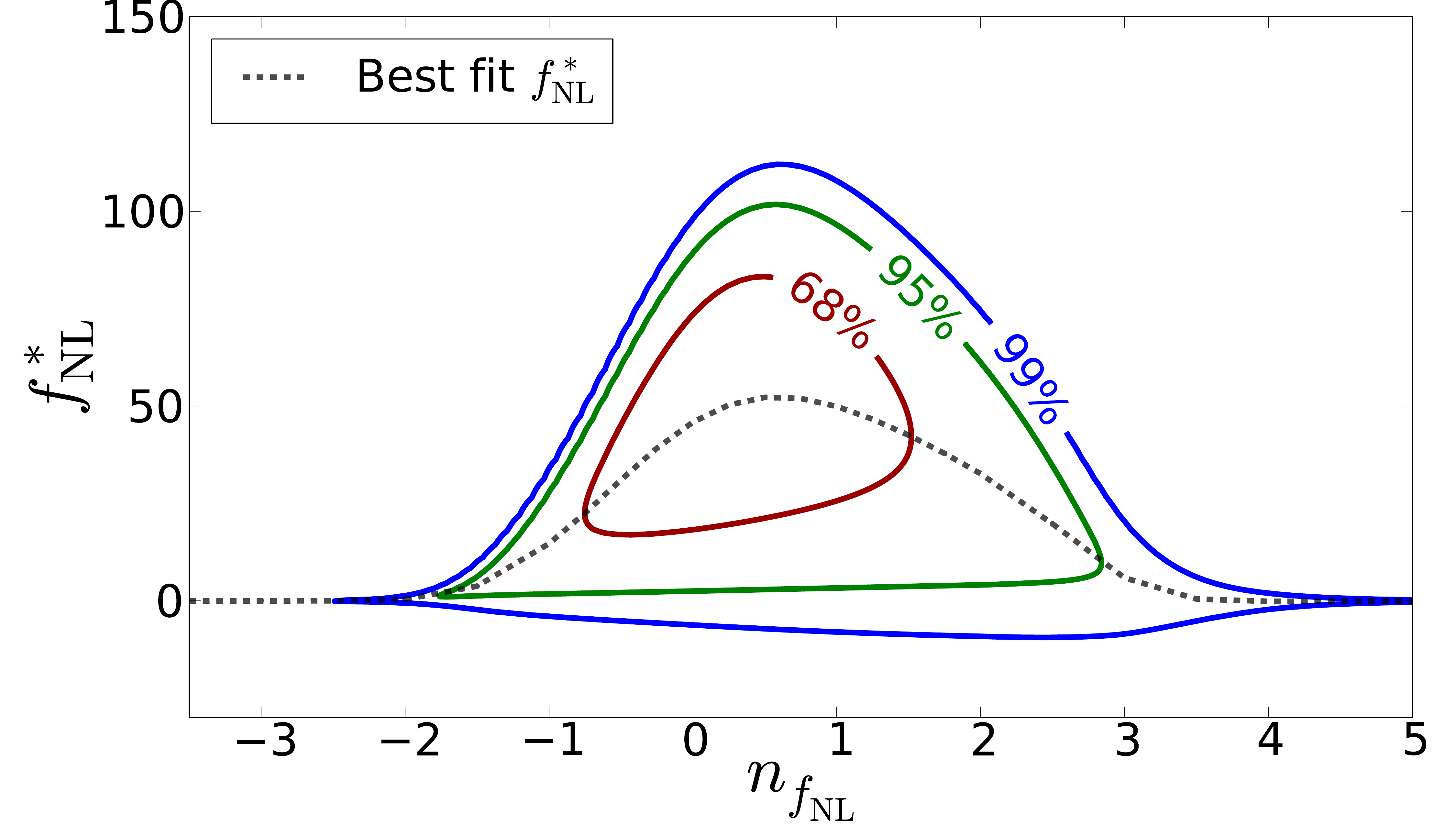}
\includegraphics[width= 3.5in]{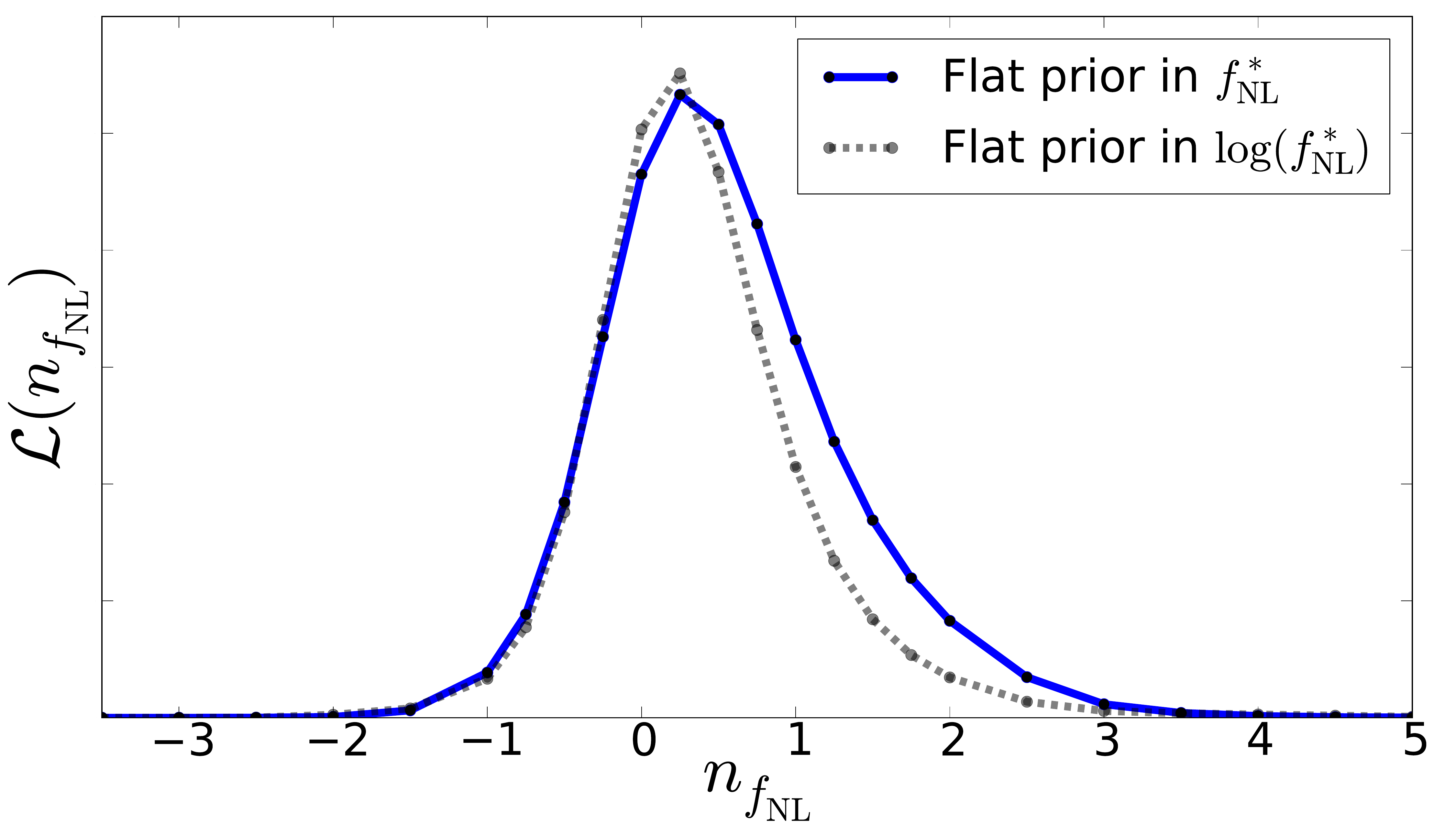}
\caption{Likelihood in the $\nf$--$\fnlstar$ plane (left panel) and
  marginalized over $\fnlstar$ (right panel).  The principal constraints,
  shown in the left panel and with the bold blue curve on the right,
  correspond to the flat prior on $\fnlstar$ at the pivot value where the
  constraints on $\fnlstar$ and $\nf$ are uncorrelated (see
  Eq.~(\ref{eq:kpiv})).  In the right panel we also show the marginalized
  likelihood for $\nf$ with a prior on $\fnlstar$ that is uniform in
  $\log(\fnlstar)$ for $|\fnlstar|>0.1$ and zero otherwise. The dashed curve
  in the left panel shows the quantity $\hat f^*_{\rm NL}$, which is the
  best-fit value of the parameter $\fnlstar$ for a fixed $\nf$. See text for
  other details. %\adam{Figures are updated!}
  }
\label{fig:likelihood}
\end{center}
\end{figure*}

A more interesting task is to calculate the
constraints when $\nf$ is allowed to vary.
With an expression for $\chi^2(\fnl^*, \nf)$ in hand, we can write an expression for the
likelihood, $\mathcal{L}(\fnlstar, \nf)\propto \exp (- {\chi^2/2})$ (dropping
the constant term with $\chi_0^2$)
\begin{align}
\mathcal{L}(\nf, \fnl^*) 
\propto  \exp\left [- {F \left( \fnl^* -  \hat{f}^*_{\rm NL} \right)^2 \over
      2}\right ] 
\exp\left [{(\fnlstarhat)^2 F \over 2}\right ]
\end{align}
To marginalize over $\fnl^*$ is also straightforward
\begin{equation}
\mathcal{L}(\nf) 
= \int \mathcal{L}(\nf, \fnl^*) \: d\fnl^* 
\propto {1 \over \sqrt{F}} \exp\left [{(\fnlstarhat)^2 F \over 2}\right ],
\end{equation}
where, recall, $F(\nf)$ is defined in Eq.~(\ref{eq:Fisher}).

\medskip
{\it WMAP7 constraints on $\nf$.} Figure \ref{fig:likelihood} shows the
likelihood $\mathcal{L}$ in the $\nf$ -- $\fnlstar$ plane, as well as the
likelihood for $\nf$ alone, calculated from the WMAP7 temperature maps.  We
used a weighted and masked combination of the WMAP V and W band maps
with the monopole and dipole subtracted, as recommended by the WMAP
team \cite{wmap5}. To extract full information from WMAP maps, we used
multipoles out to $\ellmax=800$ for the sums in Eqs.~(\ref{eq:A_map}),
(\ref{eq:B_map}) and (\ref{eq:Fisher}). We did not find a significant
improvement between $\ellmax = 700$ and $\ellmax = 800$; we chose the higher
value to be conservative in our analysis.

The quantity $\chi^2$ is independent of our choice for $k_{\rm piv}$,
but the likelihood itself is not, since $F$ is inversely proportional to
$k_{\rm piv}^{2 \nf}$.  The true pivot scale favored by the data is the value
of $k_{\rm piv}$ for which the errors in $\fnlstar$ are uncorrelated with the
errors in $\nf$. We find this scale by using the likelihood to calculate the
covariance matrix $\mathbf{C}$ between $\fnlstar$ and $\nf$
\begin{equation}
\mathbf{C}_{p_i, p_j}  = \langle (p_i - \bar{p_i})(p_j - \bar{p_j}) \rangle.
%\label{eq:covariance_nfnl}
\end{equation}
We can easily find the pivot value $k_{\rm piv}$ that diagonalizes the
covariance matrix $\mathbf{C}$ (see e.g.\ Ref.~\cite{Shandera2010})
\begin{equation}
\kpiv = k_* \exp \left( - { \mathbf{C}_{\fnlstar, \nf} \over \fnlstar \mathbf{C}_{\nf, \nf} } \right).
\label{eq:kpiv}
\end{equation}
where $k_*$ is the (arbitrary) pivot used initially, and $\fnlstar$ is the
corresponding value used in $\mathbf{C}$. Despite the fact that $k_*$ and
$\fnlstar$ show up in the expression, $\kpiv$ does not depend on them: it is a
fixed number telling us roughly where the experiment has greatest power (and
where normalization and running of $\fnl(k)$ are precisely uncorrelated).  We
find that %\adam{ Actually updated!} 
$\kpiv^{\rm WMAP7} \approx 0.064 \, \hmpcinv$. The 68\%, 95\%, and
99\% constraints on $\fnlstar$ and $\nf$ are shown at the left panel of Figure
\ref{fig:likelihood}, assuming flat priors on $\fnlstar$ and $\nf$ 
%\red{\sout{(with the weak hard-bound prior $|\nf| < 10$)}} 
and
$k_* = \kpiv^{\rm WMAP7} \approx 0.064 \, \hmpcinv$.

\medskip
{\it Dependence on the prior.}  As with most present-day cosmological
measurements, the precise constraints depend on the prior probability on the
parameters we are constraining. Even for a simple flat prior on $\fnlstar$ and
$\nf$, the actual effective prior depends on the {\it a priori} chosen pivot
in wavenumber $k_*$. For example, a flat prior on
$(\fnlstar)^{(1)}\equiv\fnl(k_{*, 1})$ defined at some pivot $k_{*, 1}$
corresponds to a non-flat prior on some $(\fnlstar)^{(2)}\equiv\fnlstar(k_{*,
  2})$ defined at some other pivot $k_{*, 2}$, since
$(\fnlstar)^{(2)}\equiv(\fnlstar)^{(1)} (k_{*, 2}/k_{*, 1})^\nf$.  If we
assume some alternate pivot $k_{*, 2}$ but hold the flat prior in $\fnlstar$,
the contours in the $\nf$--$\fnlstar$ plane (left panel of
Fig.~\ref{fig:likelihood}) are stretched vertically by a factor of
%\dragan{updated?} \adam{good catch!} 
$(k_{*,
  2}/0.064\:\hmpcinv)^\nf$.

We have experimented with different k-pivot values for a flat prior on
$\fnlstar$ and $\nf$. We have also investigated other possibilities, such as
the prior that assigns equal weight to each decade in $|\fnlstar|$ above 0.1
(so uniform in $\log(\fnlstar)$, but cut off at the arguably lowest-ever
observable value of $|\fnlstar| =0.1$ so that the total integrated likelihood is
finite).  We present the two aforementioned examples, showing constraints on
$\nf$ marginalized over $\fnlstar$, in the right panel of
Fig.~\ref{fig:likelihood}. In the end, we decide to quote results for the flat
prior and the uncorrelated $\kpiv$ value from Eq.~(\ref{eq:kpiv}), which most
closely follows priors to both non-Gaussian and other cosmological parameters
applied in the literature.

Putting it all together, we can get the  estimate for $\nf$ from the
WMAP7 data for a flat prior on $\fnlstar$ at the pivot $\kpiv$ from
Eq.~(\ref{eq:kpiv}). The $68\%$ $(95\%)$ confidence interval is %\adam{ Actually updated!}
\begin{equation}
\nf =0.30^{+0.78 \, (1.9)}_{-0.61 \, (1.2)} \,.
\label{eq:nfnl_result} 
\end{equation}
\begin{figure}[t] %htbp, h = here, t = top, b = bottom, p = page: if nothing there...
\begin{center}
\includegraphics[width= 3.5in]{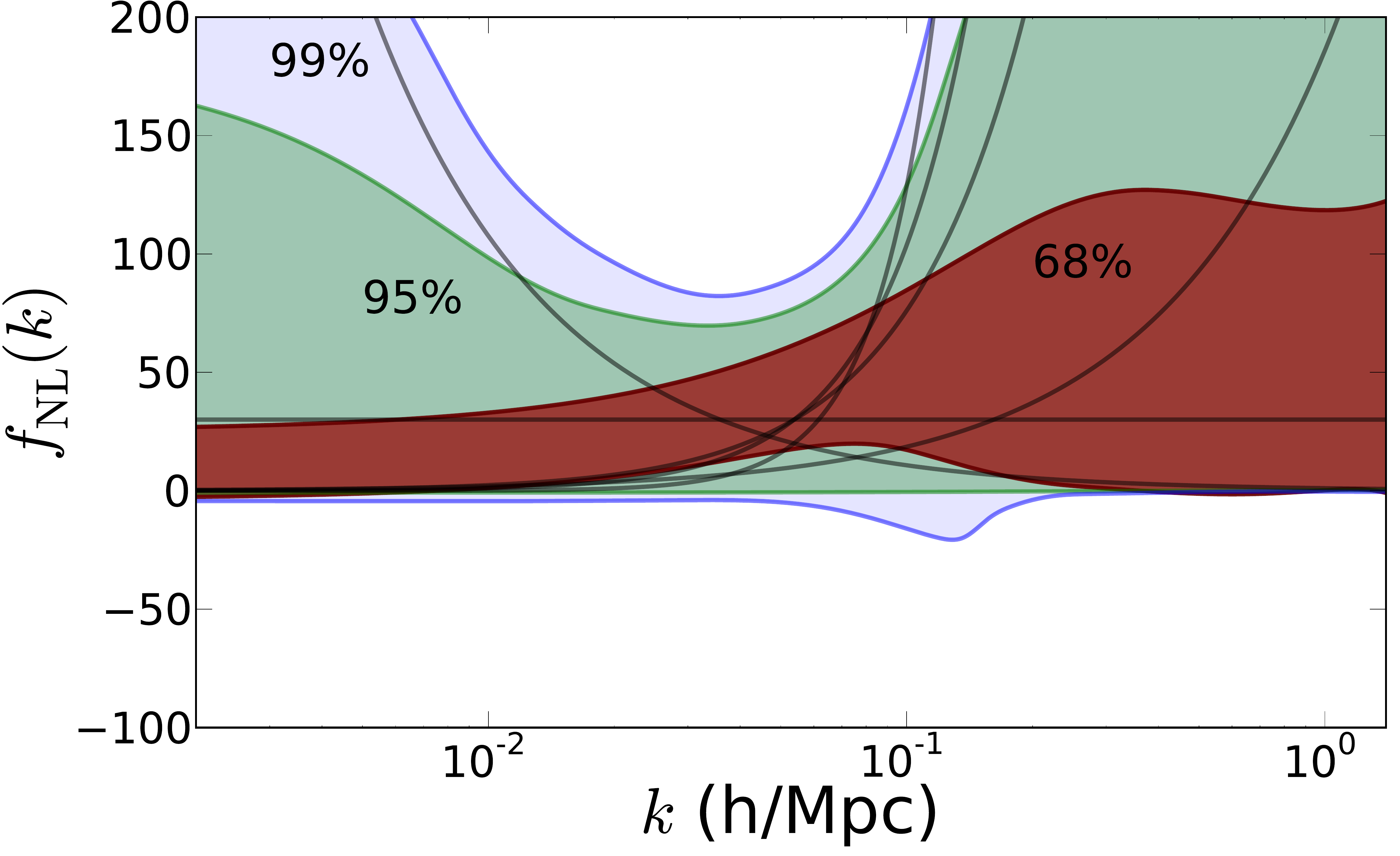}
\caption[WMAP7 constraints on $\fnl(k)$]{Constraints propagated to
  $\fnl(k)$. We also show several models that are reasonable fits to the data
  (all within the 99\% confidence limit of the left panel of
  Fig. \ref{fig:likelihood}) to guide the eye as to how typical models from
  our ansatz behave.
  }
\label{fig:errorprop_WMAP7}
\end{center}
\end{figure}
%
%\dragan{Not true any more!} \sout{Figure \ref{fig:likelihood} indicates that only a small fraction of the
%likelihood volume spills into the $\nf < 0$ region, and a value of $\fnlstar
%\approx 0$ is favored in that region. This makes physical sense: if $\fnlstar
%\neq 0$ and $\nf < 0$, $\fnl(k)$ blows up at large scales (i.e.\ small $k$),
%and we do not see evidence of this in the CMB sky.} \dragan{New sentence:}
The current constraints are therefore fully consistent with no running,
  as Fig.~\ref{fig:likelihood} clearly indicates.
Figure
\ref{fig:errorprop_WMAP7} shows the constraints in the $\fnl(k)$ plane together with a
few representative models allowed by the data.

\medskip
\textit{Conclusions.}  We have presented the first constraints on the
scale-dependence of (any form of) non-Gaussianity using the WMAP7 data. The
constraints are compatible with zero running, $\nf =0$, with very mild ($<1$-sigma)
preference for a positive value of $\nf$. We will learn more soon:
the Planck data and the data from upcoming large-scale structure surveys
should be able to improve constraints on the running of non-Gaussianity by about an order
of magnitude \cite{Sefusatti2009,Giannantonio_forecasts,Becker2012}, thus
shedding important new light on the physics of inflation.

\medskip
{\it Acknowledgements.}  We thank Kendrick Smith for initial encouragement, and
%\red{\sout{Chris Byrnes and }}\adam{I think it's safer to add this in after the paper has been officially accepted} 
Eiichiro Komatsu for useful communications. We acknowledge the use of the
publicly available CAMB \cite{CAMB} and HEALPix \cite{healpix} packages.  We
have been supported by DOE OJI grant under contract DE-FG02-95ER40899, NSF
under contract AST-0807564, and NASA under contract NNX09AC89G.

\bibliography{fnl}

\end{document}